\documentclass[%
 aip,
 jmp,%
 amsmath,amssymb,
preprint,%
author-year,%
author-numerical,%
]{revtex4-1}

\usepackage{graphicx}
\usepackage{dcolumn}
\usepackage{bm}
\usepackage[colorlinks=true,linkcolor=blue,citecolor=blue]{hyperref}

\begin{document}

\newcommand{\fr}{f_{\mathrm{0}}}
\newcommand{\VG}{V_{\mathrm{G}}}
\newcommand{\VD}{V_{\mathrm{D}}}
\newcommand{\ZG}{Z_{\mathrm{G}}}
\newcommand{\CGL}{C_{\mathrm{G1}}}
\newcommand{\CGR}{C_{\mathrm{G2}}}
\newcommand{\CQL}{C_{\mathrm{Q1}}}
\newcommand{\CQR}{C_{\mathrm{Q2}}}
\newcommand{\CQ}{C_{\mathrm{Q}}}
\newcommand{\CG}{C_{\mathrm{G}}}
\newcommand{\vF}{v_{\mathrm{F}}}
\newcommand{\nimp}{n_{\mathrm{imp}}}

\title{Contact-less characterizations of encapsulated graphene p-n junctions}

\author{V.~Ranjan}
 \email{vishal.ranjan@unibas.ch}

\author{S.~Zihlmann}
\affiliation{Department of Physics, University of Basel, Klingelbergstrasse 82, 4056 Basel, Switzerland}

\author{P.~Makk}
\affiliation{Department of Physics, University of Basel, Klingelbergstrasse 82, 4056 Basel, Switzerland}

\author{K.~Watanabe}
\affiliation{National Institute for Material Science, 1-1 Namiki, Tsukuba, 305-0044, Japan}

\author{T. Taniguchi}
\affiliation{National Institute for Material Science, 1-1 Namiki, Tsukuba, 305-0044, Japan}

\author{C.~Sch\"onenberger}
\affiliation{Department of Physics, University of Basel, Klingelbergstrasse 82, 4056 Basel, Switzerland}

\date{\today}

\begin{abstract}
Accessing intrinsic properties of a graphene device can be hindered by the influence of contact electrodes. Here, we capacitively couple graphene devices to superconducting resonant circuits and observe clear changes in the resonance- frequency and -widths originating from the internal charge dynamics of graphene. This allows us to extract the density of states and charge relaxation resistance in graphene p-n junctions without the need of electrical contacts. The presented characterizations pave a fast, sensitive and non-invasive measurement of graphene nanocircuits.

\end{abstract}


\maketitle
\section{Introduction}
In the past decade, extensive studies on graphene have unfolded interesting physics of Dirac particles on chip~\cite{Neto2009, Sarma2011, Geim2013, Liu2016}. Up to now the main technique to study the electronic properties of graphene has been low frequency lock-in technique where electrical contacts are needed  for conductance measurements. The key drawbacks of contact electrodes are highly doped regions in the vicinity of the contacts resulting in unwanted p-n junctions~\cite{Giovannetti2008} and scattering~\cite{LaMagna2011} of charge carriers. In addition, added resist residues from lithography can degrade the metal-graphene interfacial properties~\cite{Robinson2011} or even the overall device quality. An important example of this is graphene spintronics~\cite{Han2014}, where device performance is often limited by the contacts, which cause spin-relaxation and decrease of the spin-lifetime~\cite{Volmer2013,Maassen2012,Stecklein2016,Amamou2016}. Therefore, contact-less characterization, such as, microwave absorption~\cite{Obrzut2016} can open up new ways to probe inherent properties of the studied system. Here, we demonstrate such a contact-less measurement scheme by capacitively coupling graphene devices to a gigahertz resonant circuit, stub tuner~\cite{Hellmann2012}. This circuit allows us to extract both the quantum capacitance and the charge relaxation resistance with a single measurement even in the absence of electrical contacts.

We have used high mobility graphene encapsulated in hexagonal boron nitride~\cite{Wang13, Kretinin2014} which separates the graphene from external perturbations and allows local gating of the graphene flake. By forming a p-n junction the internal charge dynamics of the graphene circuit can be probed and by analyzing the microwave response of the circuit the charge relaxation resistance as well as the quantum capacitance can be inferred. Our measurements allow us to study p-n junctions in a contact-less way, which are potential building blocks of electron optical devices~\cite{Chen2016, Cheianov2007,Rickhaus2013,Grushina2013, Rickhaus2015,Rickhaus15b, Lee15}.
\begin{figure}[b!]
	\centering
		\includegraphics[width=\columnwidth]{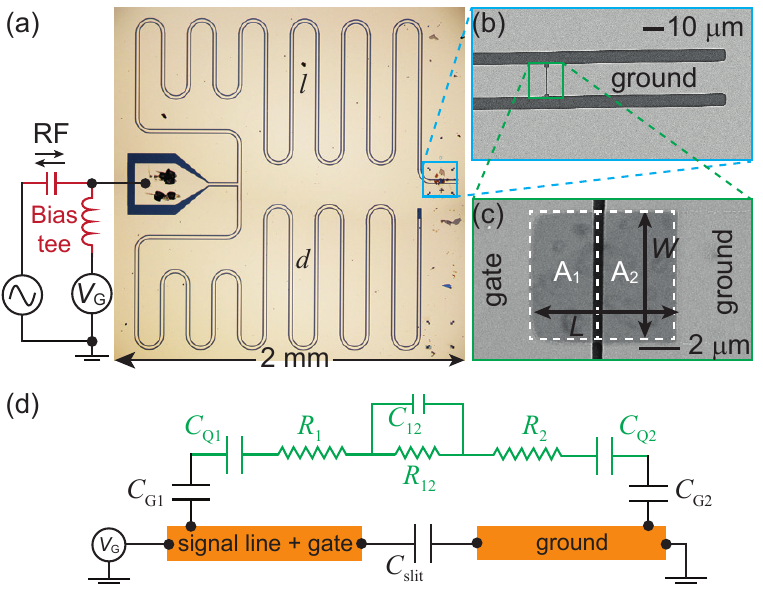}
		\caption{\label{fig:fig1}(Color online) Sample layout. (a) An optical picture of the stub tuner with arm-lengths $l$ and $d$. Central conductor and gap widths of the transmission lines are $15~\mu$m and $6~\mu$m respectively. (b) An SEM image near the $l$ end showing a narrow slit between the signal line and the ground plane (c) An SEM image of a hBN-Graphene-hBN stack for device B placed over the slit. Areas $A_1$ and $A_2$ correspond to two parts of graphene lying on the signal line and the ground plane. (d) An equivalent circuit with lumped capacitance and resistance elements. }
\end{figure}
%

\section{Device layout}
Figure~\ref{fig:fig1} shows the layout of a typical device. The stub tuner circuit is based on two transmission lines TL$_1$ and TL$_2$ of lengths $l$ and $d$, respectively, each close to $\lambda /4$~\cite{Hellmann2012}. The circuit is patterned using a 100~nm thick niobium film by e-beam lithography and subsequent dry etching with Ar/Cl$_2$. To minimize microwave losses, high resistive silicon substrates (with 170$\,$nm of SiO$_2$ on top) are used. The signal line of TL$_1$ features a slit of width $\sim$450~nm near the end before terminating in the ground plane as shown in Fig.~\ref{fig:fig1}(b,c). We place the graphene stack, encapsulated in hexagonal boron nitride (hBN), over the slit. The hBN/graphene/hBN stack is prepared using the dry transfer method described in Refs.~\onlinecite{Wang13,Zomer2014}, and positioned in the middle of the slit such that parts of the flake lie on the signal line and parts on the ground plane. We then etch the stack with SF$_6$ in a reactive ion etcher to create a well defined rectangular geometry. Some bubbles resulting from the transfer can also be seen in Fig.~\ref{fig:fig1}(c). Raman spectra are taken to confirm the single layer nature of graphene flakes (see the supplementary material).

Since there are no evaporated contacts on graphene, the same circuit can be employed for different stack geometry. We first fabricated a device with stack dimensions $W \times L$ of $6.5~\mu \mathrm{m} \times 13~\mu \mathrm{m}$ (device A), where $W$ and $L$ respectively denote the width and length of the rectangular graphene. After measurements on device A, the stack is etched into new dimensions of $6.5~\mu \mathrm{m} \times 7.2~\mu \mathrm{m}$ (device B). For both devices, a graphene area of $6.5~\mu \mathrm{m} \times 3.4~\mu \mathrm{m}$ stays on the signal (gate) line, see Fig.~\ref{fig:fig1}(c). The graphene sections lying above the ground plane had areas of $6.5~\mu \mathrm{m} \times 9.6~\mu \mathrm{m}$ for device A, and $6.5~\mu \mathrm{m} \times 3.8~\mu \mathrm{m}$ for device B. Device A is hence asymmetric while B is quasi symmetric around the slit. More importantly, two devices on the same circuit with the same graphene flake but different geometry provide consistency checks. A third symmetric device C of dimensions $5~\mu \mathrm{m} \times 12~\mu \mathrm{m}$ with a separate resonator circuit and a different graphene stack was also measured.

\section{Measurement principle}
We extract the graphene properties by measuring the complex reflection coefficient of the stub-tuner, which depends on the RF admittance of a load~\cite{Ranjan2015}. The reflected part of the RF (radio frequency) probe signal fed into the launcher port of the circuit is measured using a vector network analyzer. To tune the Fermi level of the graphene a DC voltage, $\VG$, is also applied to the launcher port with the help of a bias tee, as shown in Fig.~\ref{fig:fig1}(a). The gate voltage changes (locally) the carrier density and hence the quantum capacitance. By analyzing the response of the circuit, changes in differential  capacitance, related to the quantum capacitance $\CQ$ and in dissipation, related to charge relaxation resistance $R$ can be extracted. All reflectance measurements are performed at an input power of $-110~$dBm and at temperature of 20$\,$mK.

To understand the effect of gating, we divide the graphene into two areas denoted by $A_1$ and $A_2$ in Fig.~\ref{fig:fig1}(c). A gate voltage on the signal line induces charges on the part of graphene flake above it. Since the total number of charges in graphene in absence of a contact cannot change, charges on one part must be taken from the other. For a pristine graphene with the Fermi level at the charge neutrality point (CNP) without gating, this results in the formation of a p-n junction near the slit at each gate voltage. However, when a finite offset doping is present an offset voltage has to be applied and the charge neutrality is reached at two different gate voltages, once for each part of graphene. At voltages higher than these offset voltages (in absolute value) a p-n junction is present in the graphene. The charge carrier density changes rapidly close to the slit, but it is constant further away from the slit. Due to different areas $A_1$ and $A_2$, the applied gate voltage results into different charge densities, but equal and opposite total charge on the two sides.

In the transmission line geometry, the RF electric field emerges from the signal plane and terminate on the ground plane. While the field lines are quasi-perpendicular to the graphene surface further away from the slit, they become parallel and relatively stronger in magnitude near the slit. The field distribution hence probes both the properties of the bulk graphene (homogeneous charge distribution) and the junction graphene (inhomogeneous charge distribution). For simplicity, we model the graphene as lumped one dimensional elements of capacitance and resistance as shown in Fig.~\ref{fig:fig1}(d). The graphene impedance is then simply given as $Z_G \sim R+1/(j\omega C)$ with the total series capacitance $C$ and resistance $R$ as
\begin{align} \label{eq:RandC}
\frac{1}{C}=&\frac{1}{\CGL}+\frac{1}{\CQL}+\frac{1}{\CGR}+\frac{1}{\CQR},\\
R=&R_1+R_{12}+R_2,
\end{align}
where $\omega = 2 \pi f$ the angular frequency. Thus $C_\mathrm{Q}=\CQL\CQR/(\CQL+\CQR)$ and $C_\mathrm{G}=\CGL\CGR/(\CGL+\CGR)$ are the total quantum and geometric capacitances of the graphene device. We have assumed that the junction capacitance $C_{12}$ is relatively small so that the junction resistance $R_{12} \ll 1/(\omega C_{12})$. Moreover, we ignore the parallel slit capacitance $C_{\mathrm {slit}}$ which is small and gate independent. Together with the load $\ZG$, the reflectance response $\Gamma$ of the stub tuner can now be described by $[(Z_\mathrm{in}-Z_0)/(Z_\mathrm{in}+Z_0)]^2$ where the input impedance $Z_\mathrm{in}$ is given as~\cite{Pozar05}
\begin{equation}
Z_{\mathrm {in}} = Z_0\left( \tanh(\gamma d)+ \frac{Z_0+Z_{\mathrm {G}} \tanh(\gamma l) }{Z_{\mathrm {G}} +Z_0 \tanh(\gamma l)} \right)^{-1},
\label{eq:inputZ}
\end{equation}
with $Z_0 \sim 50~\Omega$ the characteristic impedance of the transmission line, $\gamma = \alpha+i\beta$ the propagation constant,  $\alpha$ the attenuation constant, $\beta=\sqrt{\epsilon_\mathrm{eff}}\omega/c$ the phase constant, $\epsilon_{\mathrm{eff}}$ the effective dielectric constant and $c$ the speed of light.

\section{Experimental results}
\begin{figure}[t!]
	\centering
		\includegraphics[width=\columnwidth]{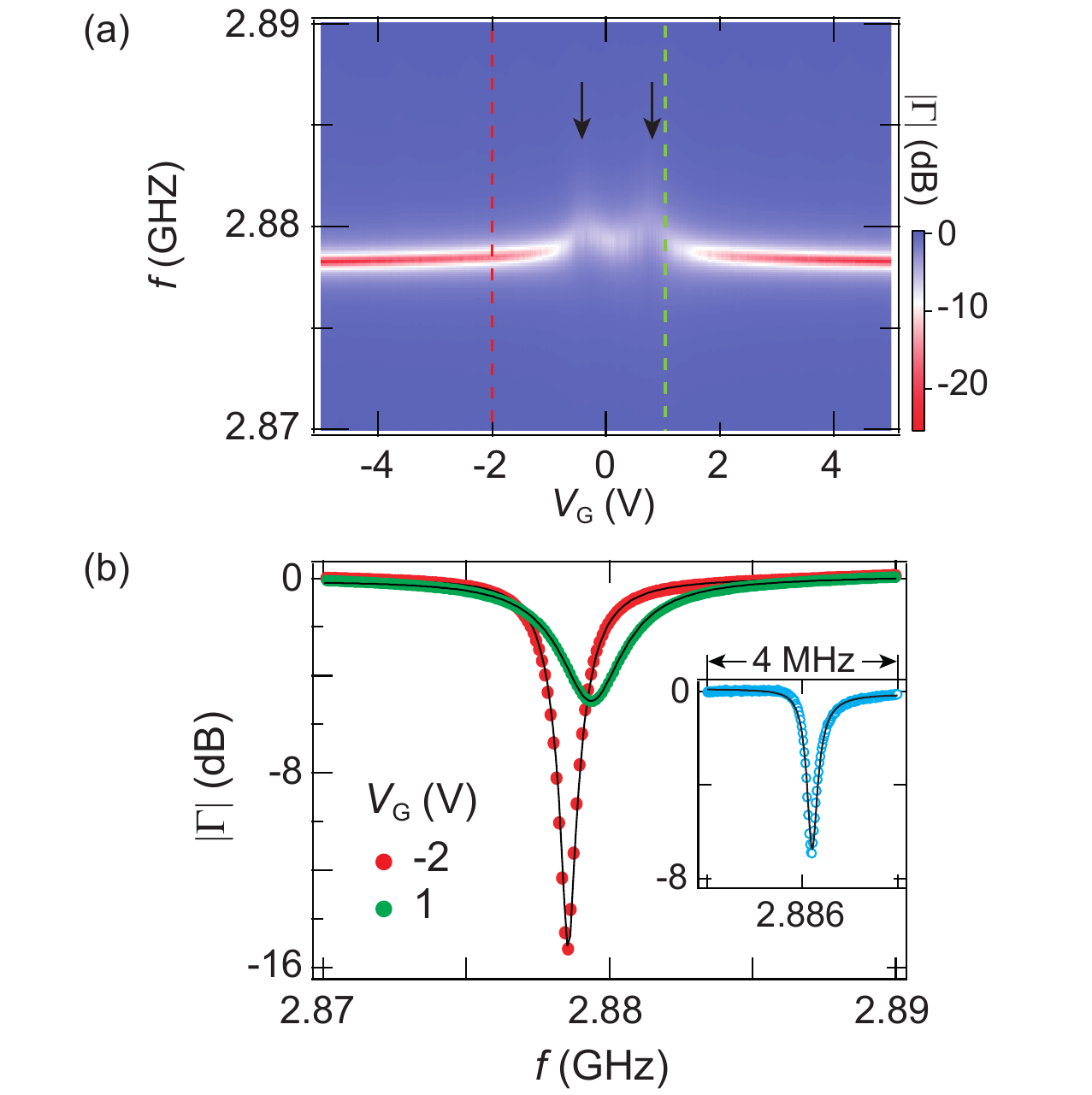}
		\caption{(Color online) Reflectance response of the stub tuner. (a) A color map of the measured reflectance power near the resonance frequency versus different gate voltages. Arrows denote the charge neutrality points (CNP). Its asymmetric separation around the zero voltage is due to inhomogeneous distribution of an average offset doping $\sim 3\times 10^{11}$~cm$^{-2}$ in the system. (b) Main panel: Cuts of the reflectance curves at two different gate voltages with fits to the Eq.~\ref{eq:inputZ}. Inset: The reflectance response of the same device without graphene. The input RF power is $-110~$dBm which corresponds to an AC excitation amplitude of $0.7~\mu$V. }
	\label{fig:fig2}
\end{figure}
Figure~\ref{fig:fig2}(a) shows a color map of frequency and gate voltage response of the reflected signal for device B. Large frequency shifts at two gate voltages can be observed near $\VG=0$. These can be identified as points where either part of the graphene flake is driven charge neutral. At higher gate voltages, p-n junctions are formed in between the unipolar regimes. This behaviour is observed in all our devices, suggesting the presence of a finite offset doping in the system. From the vertical cuts of the map shown in Fig.~\ref{fig:fig2}(b), changes in the resonance-depth, -width and -frequency are apparent. Naively, a pure capacitive load should shift the resonance frequency, while a pure resistive load changes dissipation of the system.

\begin{figure}[t!]
	\centering
		\includegraphics[width=\columnwidth]{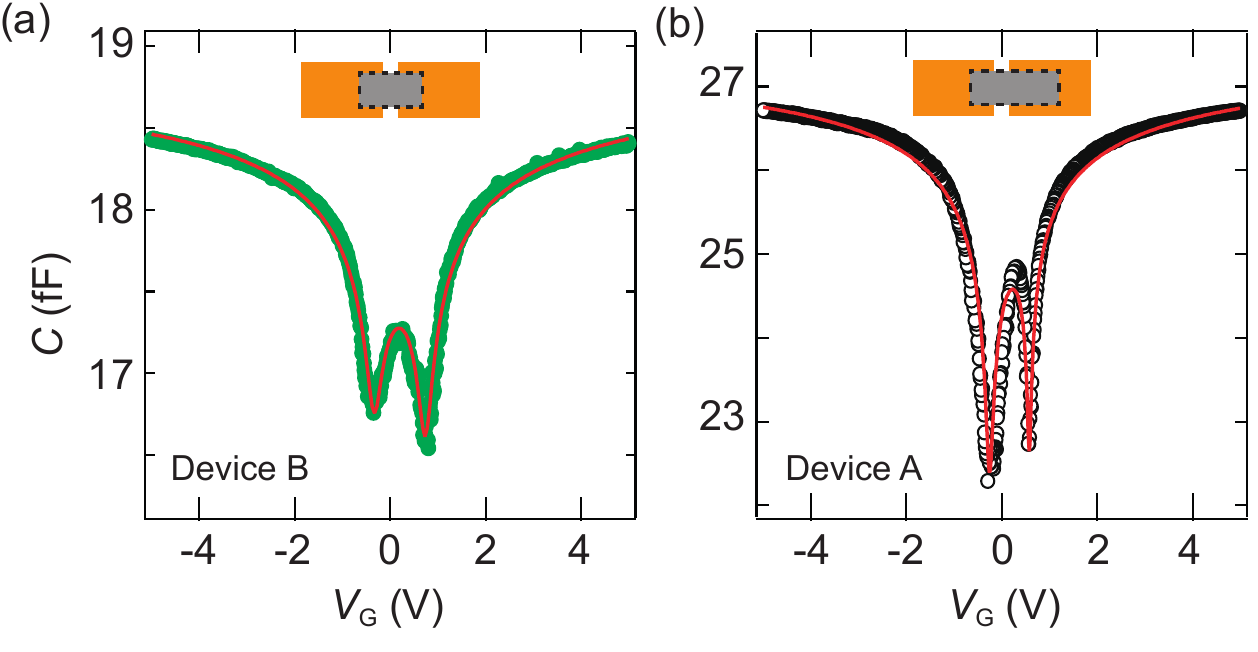}
		\caption{Quantum capacitance of graphene (a) The extracted capacitance from the fitting of the reflectance response to Eq.~\ref{eq:inputZ} for device B and (b) for device A. Error bars are smaller than the symbol size. Solid lines are the best fits to Eq.~\ref{eq:RandC} showing good agreement with the graphene density of states. Insets: schematics of relative dimensions of graphene flake across the slit.}
	\label{fig:fig3}
\end{figure}

\begin{figure}[t!]
	\centering
		\includegraphics[width=\columnwidth]{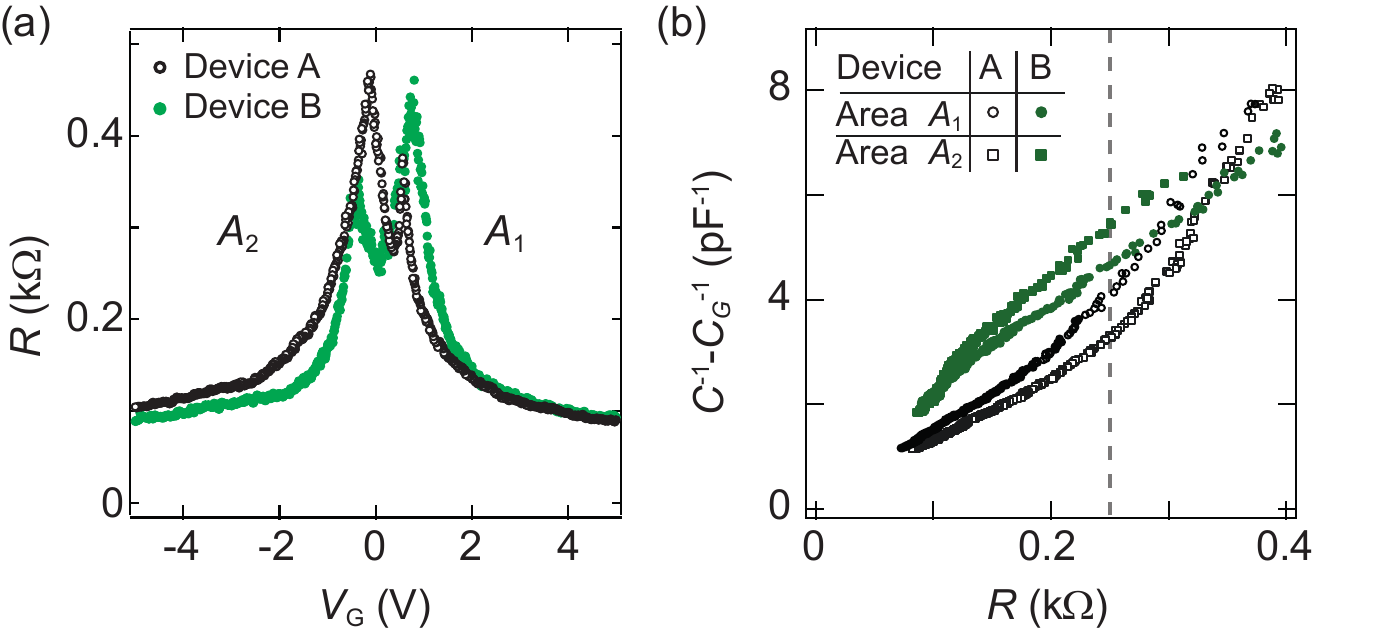}
		\caption{(Color online) Dissipation in graphene (a) The extracted charge relaxation resistance for two devices fabricated on the same hBN-graphene-hBN stack. The same loss constant is used in fitting the reflectance map. (b) Inverse quantum capacitance, obtained by subtracting geometric capacitance from the total extracted capacitance, as a function of the simultaneously extracted charge relaxation resistance.}
	\label{fig:fig4}
\end{figure}

To quantitatively extract $\ZG$, we first need to extract the parameters $l$, $d$, $\alpha$ and $\epsilon_{\mathrm{eff}}$ from the reflectance measurements of the same circuit without any graphene stack. To this end, we simply ash the graphene stack away using Ar/O$_2$ plasma. The frequency response of the open circuit is shown in the inset of Fig.~\ref{fig:fig2}(b) together with a fit to Eq.~\ref{eq:inputZ} with $\ZG=\infty$. We extract $l \approx 10.57\,$mm and $d \approx 10.39\,$mm, $\alpha \approx 0.0025\,$m$^{-1}$ and the effective dielectric constant $\epsilon_{\mathrm{eff}}\approx 6.1$. The loss constant corresponds to an internal quality factor of 25,000 which is readily achieved with superconducting Nb circuits. The extracted lengths are within $1\%$ of the designed geometric lengths. Moreover, the resonance frequency of the open stub tuner (2.886~GHz) is larger than the values observed in Fig.~\ref{fig:fig2}(a), confirming the capacitive load of our devices. We now fix the extracted parameters from open circuit, and fit the resonance spectra to deduce $R$ and $C$. As shown in Fig.~\ref{fig:fig2}(b), the fitting to Eq.~\ref{eq:inputZ} yields $R=118\,\Omega$, $C=18.2\,$fF for $\VG=-2\,$V and $R=328\,\Omega$ and $C=17.2\,$fF for $\VG=1\,$V. Similar fitting is performed at all gate voltages and deduced $C$ and $R$ are plotted in Fig.~\ref{fig:fig3} and \ref{fig:fig4}.

As shown in Fig.~\ref{fig:fig3}, we observe for both devices a double dip feature in the extracted capacitance near $\VG=0\,$V and its saturation at higher voltages. While the dips have similar widths for device B, these are quite different for device A. This again results from the asymmetric gating of the two areas of graphene. To understand the general dependence, we look back at the individual capacitance contributions in Eq.~\ref{eq:RandC}. Geometric capacitance $C_{\mathrm Gi}$ with $i=1,2$ is simply given by $C_{\mathrm Gi}=A_i\epsilon_0 \epsilon_\mathrm{BN}/d$, where $\epsilon_0$ is the vacuum permittivity, $\epsilon_\mathrm{BN}$ the dielectric constant, and $d=21.5~$nm the thickness of the bottom hBN estimated from AFM measurements. Additionally, the quantum capacitance can be derived from the density of states~(DoS) as $C_\mathrm{Q}/A=e^2 \cdot$DoS. The resulting dependence of $C_\mathrm{Q}$ in graphene with gate voltage $V$ is then explicitly given as~\cite{Chen2008, Xia2009, Droscher2010,Yu2013}
\begin{equation} \label{eq:CQ}
C_{\mathrm{Q}i} (V) = A_i\frac{4 e^2}{h v_\mathrm{F}}\sqrt{n_i(V)\pi},
\end{equation}
with $i=1,2$ and $v_F$ the Fermi velocity and $h$ the Planck constant. The gate induced carier density is $n_i(V)= (V_i-V^0_i)C_\mathrm{Gi}/(A_i e)$, where $V^0_i$ accounts for the offset in CNP from zero. Using Eqs.~\ref{eq:RandC} and \ref{eq:CQ}, it can be seen that the $C$ is dominated by the $C_\mathrm{G}$ at large gate voltages causing the saturation of the extracted capacitance. The saturation values are different for the two devices because different flake areas yield different $\CG$. In contrast, near charge neutrality, $\CQ \lesssim \CG$, the quantum capacitance starts to dominate. The fact that $C$ does not approach zero can be attributed to the impurity induced doping $\langle n_\mathrm{imp,i}^2\rangle$, with $i=1,2$, resulting from charge puddles~\cite{Xue2011}. To this end, we replace $n_i(V)$ with a total carrier density including this factor: $\sqrt{n_i^2(V)+n_\mathrm{imp,i}^2}$. The knowledge of most of the relevant parameters allows us to fit the capacitance curves with $\epsilon_\mathrm{BN}$, $n_\mathrm{imp}$ and $v_\mathrm{F}$ as fitting parameters. This is shown by solid curves in Fig.~\ref{fig:fig3}. The excellent fits to Eq.~\ref{eq:RandC} capture both the depth and width near the Dirac charge neutrality points and justify the series model of the graphene impedance with $C$ arising from the total graphene area. For device A(B), we extract $\epsilon_\mathrm{BN} \approx 4~(4)$, $\vF \approx 1.05~(0.95) \times 10^6~$m/s and $n_\mathrm{imp,1}\approx 5~(7)\times10^{10}$~cm$^{-2}$ and $n_\mathrm{imp,2} \approx 1~(6)\times10^{10}$~cm$^{-2}$. The low impurity carrier concentration is consistent with transport measurements in graphene encapsulated with hBN~\cite{Xue2011}. In another symmetric device C (see the supplementary material) with a different circuit and a different stack, the $\nimp$ is found to be even lower $\approx 4\times 10^{9}~$cm$^{-2}$ and extracted Fermi velocity higher $\approx 1.54 \times 10^6~$m/s. Such renormalization of $\vF$ due to electron-electron interactions at low doping has been observed both in capacitance~\cite{Yu2013} and transport measurements~\cite{Ponomarenko2010, Elias2011, Chae2012} in homogeneously doped graphene.

We now discuss the real part of the graphene impedance which relates to the dissipation of the microwave resonance. The extracted $R$ for two devices fabricated from the same hBN-graphene-hBN stack (device A and B) is plotted in Fig.~\ref{fig:fig4}(a). Two peaks are visible in the extracted resistances, which are similar to the charge neutrality points in transport measurements. The positions of the peaks correspond to the minima of the extracted capacitance. At large gate voltages where residual impurities play a negligible role, the resistances start to saturate around similar values despite the fact that device A is twice as long as device B. In the absence of contacts, this points to the direction that the resistance is dominated by the p-n junction at high doping. A similar behaviour of $R$ is seen in the device C (see the supplementary material). Close to CNPs, the respective bulk graphene areas also contribute significantly to the resistance. These features are in agreement with the carrier density ($n$) dependence of the bulk and p-n junction conductivity in graphene. While the conductivity for the p-n junction~\cite{Cheianov2006} is proportional to $n^{1/4}$, it scales as $n$ or $n^{1/2}$ for bulk graphene depending on the relevant scattering mechanisms~\cite{Pallecchi2011}. 

The bulk carrier transport in graphene can be characterized by the diffusion constant $D$. By knowing both $R$ and $\CQ$, $D$ can be calculated from the Einstein relation
\begin{equation} \label{eq:Einstein}
 D= (L)^2/(R\CQ).
\end{equation} 
A complication in our devices arises due to the presence of p-n junction which is almost always present. We can, therefore, only get an estimate of $D$ by considering $R$ and $\CQ$, that are largely arising from only one graphene area $A_1$ or $A_2$. For higher gate voltages, the p-n junction resistance plays a role, whereas close to the CNPs, both areas  contribute to the resistance and the capacitance significantly. The inverse of the quantum capacitance, obtained by subtracting the total geometric contribution $C_\mathrm{G}$ from the total extracted $C$, is now plotted against the simultaneously measured resistance $R$ in Fig.~\ref{fig:fig4}(b). We have taken the data points that are strictly on the left (negative $\VG$) or the right side of CNPs (positive $\VG$). We extract $D$ at a modest doping marked by the dashed line in Fig.~\ref{fig:fig4}(b). Since one cannot separate the contribution of p-n resistance, by using the total $R$ in Eq.~\ref{eq:Einstein}, bulk graphene resistance is overestimated and therefore $D$ is underestimated. In graphene areas $A_1$ lying on signal plane (not changed after etching), we get $D=$0.19~(0.21)$~\times 10^4$~cm$^2$/s for device A(B). In contrast for area $A_2$ lying on ground plane, we get 1.2~(0.32)$~\times 10^4$~cm$^2$/s. The large differences in $D$ for area $A_2$ between two devices is consistent with variations in the impurity concentration extracted from fitting of the capacitance, and could result from the additional etching step of the stack for device B. We furthermore estimate an average mean free path of two areas $l_m=2\langle D \rangle/\vF$ to yield 1.4~(0.5)~$\mu$m for device~A(B), which are in reasonable agreement with values reported in transport measurements.

\section{Discussions}
In summary, we have capacitively coupled encapsulated graphene devices to high quality microwave resonators and observed clear changes in the resonance-linewidth and -frequency as a response to change in the gate voltage. We are able to reliably extract geometrical and quantum capacitance in good agreement with the density of states of graphene and simple capacitance models, respectively. Moreover, the charge relaxation resistance can be simultaneously inferred and the diffusion constant can be estimated. The results highlight fast characterizations of graphene without requiring any contacts that could compromise the device quality.

An uncertainty of the given measurements lies in the extracted $R$ due to the loss constant of the circuit which can vary from one cool down of the device to the next. From fitting the reflectance response with a different $\alpha$, we find that the extracted $R$ at different circuit losses are merely offset to each other however the extracted $C$ is not affected. The behaviour can be understood by replacing the loss constant with a resistor $R_\mathrm{Loss}$ in series with the graphene resistor $R$. The $\alpha$ could be accurately separated in quantum Hall regime, where the conductance of the device is known. For this, due to the large B-fields copper resonators~\cite{Rahim2016} have to be fabricated. The ability of our circuit to measure quantum capacitance and resistance in a contact-free way can for example be useful to study band modification of graphene due to proximity spin orbit effects~\cite{Martin2016} or due to Moire superlattices \cite{Yu2014}. The method can also be useful for other 2D materials, on which an ohmic contact is challenging to obtain.\\

\textbf{ACKNOWLEDGMENTS}\\
This work was funded by the Swiss National Science Foundation, the Swiss Nanoscience Institute, the Swiss NCCR QSIT, the ERC Advanced Investigator Grant QUEST, iSpinText Flag-ERA network, and the EU flagship project graphene. Growth of hexagonal boron nitride crystals was supported by the Elemental Strategy Initiative conducted by the MEXT, Japan and JSPS KAKENHI Grant Numbers JP26248061,JP15K21722, and JP25106006. The  authors  thank  Gerg\"{o} F\"{u}l\"{o}p for  fruitful discussions.\\

\bibliographystyle{apsrev4-1}
\bibliography{references_gr}

\end{document}